\documentstyle[prl,aps,preprint,tighten]{revtex}

\begin{document}

\title{An Exactly Solvable Kondo Problem for Interacting One-Dimensional Fermions}
\author{Yupeng Wang$^{1,2}$   and   Johannes Voit$^2$}
\address{1. Cryogenic Laboratory, Chinese Academy of Sciences, Beijing 100080, 
P. R. China\\
2. Physikalisches Institut,  Universit\"{a}t Bayreuth, D-95440 Bayreuth, Germany}%
\date{\today}
\maketitle
\begin{abstract}%
The single impurity Kondo problem in the one-dimensional (1D) $\delta$-potential 
Fermi gas is exactly solved for two sets of special coupling 
constants via Bethe ansatz. It is found that ferromagnetic Kondo screening does 
occur
in one case which confirms the Furusaki-Nagaosa conjecture 
while in the other case it 
does not, which we explain in a simple physical picture. 
The surface energy, the low temperature specific heat and the Pauli susceptibility
induced by the impurity and thereby the Kondo temperature are derived explicitly.
\end{abstract}
\newpage
With the development of nanofabrication techniques for quantum wires and
the prediction of edge states in the
quantum Hall effect, the interest in 1D electron systems has been renewed in the 
recent years.
Such systems differ fundamentally from those in three dimensions, where the low 
energy
properties can be described very well by the Landau Fermi liquid theory. In the 
presence of 
repulsive interaction, quasiparticle excitations are replaced in one dimension 
by collective excitations.\cite{1,2}
The low temperature properties of such systems are then described by the so-called
Luttinger liquid theory.
\par
Considerable attention has been focused on the response of a Luttinger liquid to 
localized perturbations. An example is the exchange
interaction between such a non-Fermi liquid and a localized magnetic impurity 
(Kondo 
problem in a Luttinger liquid). This problem was 
first considered
by Lee and Toner,\cite{3} who used abelian bosonization to map the problem onto
a kink-gas action. The dependence of the Kondo temperature $T_K$ on the bare 
exchange $J$ generically has a power-law dependence on the Kondo coupling $J$
determined by the effective Luttinger coupling $K_{\rho}$, and crosses over 
to the familiar exponential form for $J$ large or $K_{\rho} \rightarrow 1$. 
Subsequently,
a poor man's scaling treatment on this problem was performed by 
Furusaki and Nagaosa.\cite{4} They  proposed the
interesting conjecture that a Luttinger liquid supports a Kondo
effect even if the bare exchange interaction is ferromagnetic. Moreover,
they showed that the excess specific heat and Pauli susceptibility due to
the Kondo impurity are different from those of a local Fermi liquid. 

There are a few exact results on the Kondo effect in a Luttinger liquid but no
exact solution has been proposed to date. In the absence of electron-electron
interaction (Luttinger gas), the problem can be mapped onto a standard
2-channel Kondo problem\cite{wang2,12} but it is not clear, how these results
are changed for a Luttinger \em liquid. \rm A somewhat artificial model, where
the propagation direction of electrons is correlated with their spin direction,
does show both ferromagnetic and antiferromagnetic Kondo effects but its 
thermodynamics is Fermi-liquid-like and apparently differs from the 
Furusaki-Nagaosa
solution.\cite{sching} Finally, boundary conformal field theory has allowed the
classification of all possibilities of critical behavior for a Luttinger liquid
coupled to a Kondo impurity.\cite{froejdh} It turns out that there are just two
possibilities, a local Fermi liquid with standard low-temperature thermodynamics
or a non-Fermi liquid with the anomalous scaling observed by Furusaki and Nagaosa.
In this letter, we propose an exact solution to two cases of the Kondo problem 
in a Luttinger liquid. In one case, we find a ferromagnetic Kondo effect
similar to Furusaki and Nagaosa\cite{4} but
with the thermodynamics of a local Fermi-liquid while in the other case, we do not
find complete Kondo screening. 
\par
The proper starting point to consider the Kondo problem
in a fermion chain is the Hamiltonian
\begin{eqnarray}
H & = & H_0+H_{\rm imp}, \nonumber\\
\label{hubbard}
H_0 & = & -t\sum_{j=1,\sigma}^{N-1}(C_{j\sigma}^\dagger C_{j+1\sigma}+h.c.)
+U\sum_{j=1}^Nn_{j,\uparrow}n_{j,\downarrow},\\
H_{\rm imp} & = & J\sum_{\sigma,\sigma'}[C_{1\sigma}^\dagger C_{1\sigma'}
+C_{N\sigma}^\dagger C_{N\sigma'}]
{\bf \tau}_{\sigma\sigma'}\cdot{\bf S}
+V\sum_\sigma[n_{1\sigma}+n_{N\sigma}],\nonumber
\end{eqnarray}
where $H_0$ is the Hamiltonian of the Hubbard chain with open boundaries 
and $H_{imp}$ is the interaction between the electron gas and the local 
impurity;
${\bf \tau}$ is the Pauli matrix and ${\bf S}$ is the local spin-$\frac 12$ 
operator. 
However, even such a simple model can not be solved exactly. To give some exact 
information of the Kondo effect 
in a Luttinger liquid, we consider a related continuum model\cite{yang}
\begin{eqnarray}
\label{delta}
H=-\sum_{j=1}^N\frac{\partial^2}{\partial x_j^2}+2c\sum_{i<j}\delta(x_i-x_j)
+\sum_{j=1}^N
[\delta(x_j)+\delta(L-x_j)][J{\bf \tau}_j\cdot{\bf S}+V],
\end{eqnarray}
where $c>0$ is the electron-electron coupling constant; $J$ and $V$ describe the 
Kondo coupling constant and the boundary potential
respectively; $L$ is the length of the system and $N$ is the number of 
electrons. Of course, one would like to solve (\ref{hubbard}) or (\ref{delta})
with periodic boundary conditions. However, as pointed out earlier
\cite{4,5} impurities in a Luttinger 
liquid behave always like
strong scattering centers at low energy scales and effectively enforce open
boundaries. 

If a boundary is closed simply by an infinite wall, 
any electron impinging on this
boundary will be completely reflected and suffer a phase shift between
incident and reflected waves of $\pi$\cite{7,8,wang}. If, however, the wall
is preceded by a very narrow potential well, (boundary fields in the lattice
models\cite{10,alca}), the incident and reflected waves will not be exactly
in opposite phase. We therefore specify our model by imposing the boundary
conditions that the waves arriving at either end are reflected as
\begin{eqnarray}
\label{refl}
e^{i k_j x} & \rightarrow & R_{j0}(k_j) e^{-ik_jx} \;, \;\;\; (x=0) \; , \\
e^{i k_j x} & \rightarrow & R_{j0}(k_j) e^{-ik_jx - 2 i k_j L} 
\;, \;\;\; (x=L) \; ,\nonumber
\end{eqnarray}
where $R_{j0}(k_j)$ is the reflection matrix of the electrons at the boundary. 
In our case, $R_{j0}(k_j)$ is an operator rather than a $c$-number because the
electron-boundary scattering involves a spin-exchange process between the
electrons and the local impurity. The Yang-Baxter equation
\begin{equation}
\label{yb}
S_{jl}(k_j-k_l) R_{j0}(k_j) R_{l0}(k_l) = R_{l0}(k_l) R_{j0}(k_j) S_{jl}(k_j-k_l)
\end{equation}
constrains the integrability of our model. $S_{jl}(k_j-k_l)$ is the 
electron-electron scattering matrix in the bulk.
Manipulating these equations, we
find that the Hamiltonian (\ref{delta}) can be solved exactly by including an 
irrelevant local counter term proportional to 
$\sum_{j=1}^N[\delta'(x_j)-\delta'(x_j-L)]$ (ensuring that the wave function
vanishes outside the domain $0 \leq x \leq L$), and 
for the following sets of parameters:
(i) $J=-2V=-c/ 2$; (ii) $J=2V/3=-c/2$; (iii)  $J=0$. Case (iii) is just the open
boundary problem with boundary potentials of the $\delta$-potential 
Fermi gas model
which has been discussed by Woynarovich\cite{6},
and we shall not repeat the discussion here. Case (i) corresponds to a repulsive
boundary potential which, following Luttinger liquid theory, 
would not influence the occurrence of the 
antiferromagnetic Kondo effect\cite{12}. Nothing is known about its influence
on the ferromagnetic Kondo effect. Case (ii) represents an attractive boundary
potential which has not been considered previously.
\par
The eigenvalue problem of Hamiltonian (\ref{delta}) 
is similar to those of other integrable models\cite{7,8,wang,6}
and gives the following Bethe ansatz equations
\begin{eqnarray}
e^{2ik_jL}=s^2(k_j)\prod_{\alpha=1}^M
\frac{k_j-\lambda_\alpha+
i\frac{c}2}{k_j-\lambda_\alpha-i\frac{c}2}
\frac{k_j+\lambda_\alpha+i\frac{c}2}{k_j+\lambda_\alpha-i\frac{c}2},\nonumber\\
\label{ba}
\prod_{j=1}^N\frac{\lambda_\alpha-k_j+
i\frac{c}2}{\lambda_\alpha-k_j-i\frac{c}2} 
\frac{\lambda_\alpha+k_j
+i\frac{c}2}{\lambda_\alpha+k_j-i\frac{c}2}
\left(\frac{\lambda_\alpha+i\frac{c}2} {\lambda_\alpha-i\frac{c}2} \right)^2
=\prod_{\beta\neq\alpha}\frac{\lambda_\alpha-\lambda_\beta+ic}
{\lambda_\alpha-\lambda_\beta-ic}\frac{\lambda_\alpha+\lambda_\beta+ic}
{\lambda_\alpha+\lambda_\beta-ic},
\end{eqnarray}
with the eigenvalue of the Hamiltonian (\ref{delta}) as 
\begin{eqnarray}
E=\sum_{j=1}^N k_j^2,
\end{eqnarray}
where $k_j$'s and $\lambda_\alpha$'s are the ``rapidities" of the charge 
and spin respectively,
$s(k)=1$ for case (i) and $s(k)=\frac{k-ic}{k+ic}$ for case (ii); $M$ is the 
number of down spins. For convenience, we put $N$ odd
in the following discussion. Our main conclusion is that 
for case (i), the ground state of the system is a spin singlet which
supports the Furusaki-Nagaosa conjecture, while for case (ii), the ground state
of the system is a spin triplet which indicates that the local spin can not be 
screened.
\par
Case (i): No bound state exists and the ground state is described by two sets 
of real parameters $\{k\}$ and $\{\lambda\}$.
Define the quantities
\begin{eqnarray}
Z_L^c(k_j) & = & \frac 1\pi \left\{k_j - \frac 1{2L}\theta\left(k_j,\frac c2\right)
+ \frac1{2L} \sum_{\alpha=-M}^M
\theta\left(k_j-\lambda_\alpha,\frac c2\right)\right\}, \\
Z_L^s(\lambda_{\alpha}) & = & \frac 1\pi
\left\{ \frac 1{2L} \left[ 2 \theta\left(\lambda_{\alpha},
\frac c2\right)+\theta(\lambda_{\alpha},c) \right] 
+ \frac 1{2L}\sum_{j=-N}^N \theta\left(\lambda_{\alpha}-k_j,\frac c2\right)
-\frac 1{2L}\sum_{\beta=-M}^M\theta(\lambda_{\alpha}-\lambda_\beta,c)
\right\}, \nonumber
\end{eqnarray}
where $\theta(x,c)=2 \tan^{-1}x/c$ and we have used the reflection
symmetry of the Bethe Ansatz to include solutions with negative $k_{-j}=-k_j$ and
$\lambda_{-\alpha}=
-\lambda_\alpha$.
The Bethe ansatz (\ref{ba}) is solved by $Z_L^c(k_j)=\frac {I_j}L$ and 
$Z_L^s(\lambda_\alpha)=\frac {J_\alpha}L$,
where $I_j$ and $J_\alpha$ are nonzero integers. In the ground state, $I_j$ and 
$J_\alpha$
must be consecutive integers to minimize the energy. The roots $k_j$ and 
$\lambda_{\alpha}$ become dense in the thermodynamic limit, and we define their
densities as
\begin{eqnarray}
\label{dens}
\rho_L^c(k)=\frac{d Z_L^c(k)}{d k},{~~~~~~~~~~~~~~~~~}
\rho_L^s(\lambda)=\frac{d Z_L^s(\lambda)}{d \lambda}
\end{eqnarray}
for length $L$. 
Their finite size corrections are\cite{10,9}
\begin{eqnarray}
\rho_L^c(k)=\rho_c(k)+
\frac 1L\delta \rho_c(k)+O(\frac 1{L^2}),\nonumber\\
\label{finite}
\rho_L^s(\lambda)=\rho_s(\lambda)+\frac 1L
\delta \rho_s(\lambda)+O(\frac 1{L^2}).
\end{eqnarray}
The densities $\rho_c(k) $ and $\rho_s(\lambda)$ in the infinite system limit
follow from the integral equations
\begin{eqnarray} 
\rho_c(k) & = & \frac{1}{\pi} +  \int_{-\Lambda}^{\Lambda}
K \left(k - \lambda, \frac{c}{2} \right) \rho_s(\lambda) d\lambda \\
\rho_s(\lambda) & = & \int_{-k_0}^{k_0} K \left( \lambda - k, \frac{c}{2} \right)
\rho_c(k) dk - \int_{\Lambda}^{\Lambda} K ( \lambda - \lambda', c) 
\rho_s(\lambda') d\lambda' \;, \nonumber
\end{eqnarray} 
and the finite-size corrections $\delta \rho_{c}(k)$, $\delta \rho_{s}(\lambda)$
are solutions of 
\begin{eqnarray}
\delta \rho_c(k) & = & - K\left(k,\frac c2\right) + 
\int_{-\Lambda}^{\Lambda} K \left(k-\lambda,\frac c2\right)
\delta \rho_s(\lambda)d\lambda,\nonumber\\
\delta \rho_s(\lambda) & = & 2 K\left(\lambda,\frac c2\right)+K(\lambda,c)+
\int_{-k_0}^{k_0}K\left(\lambda-k,\frac c2\right)\delta \rho_c(k)dk-
\int_{-\Lambda}^{\Lambda} K(\lambda-\lambda',c)\delta \rho_s(\lambda')d\lambda',
\end{eqnarray} 
with $K(x,c)=\pi^{-1}c/({x^2+c^2})$; $k_0$ and $\Lambda$ are cutoffs of $k$ 
and $\lambda$, respectively, for the ground state, and satisfy 
$\int_{-k_0}^{k_0} \rho_L^c(k) dk = (N+1/2)/L$ and $\int_{-\Lambda}^{\Lambda} 
\rho_L^s(\lambda) d\lambda = (M+1/2)/L$. 
Notice that $\Lambda \rightarrow \infty$ when $L \rightarrow \infty$.
Finally, one can also derive similar integral equations for the dressed energies 
for the finite ($\epsilon_L^{c}(k)$, $\epsilon_L^{s}(\lambda)$) and infinite 
system\cite{9}. 
Their finite size corrections
behave as
\begin{eqnarray}
\epsilon_L^c(k) = \epsilon_c(k) + O \left(\frac{1}{L^2} \right) \;, \nonumber \\
\epsilon_L^s(\lambda) = \epsilon_s(\lambda) + O \left(\frac{1}{L^2} \right) \;,
\end{eqnarray}
where $\epsilon_c(k)$ and $\epsilon_s(\lambda)$ are the dressed energies for the
infinite system.
\par
We now discuss the physical properties of the system by evalutating its
thermodynamics. The boundary energy induced by the impurity is
\begin{eqnarray}
f=\int_{-k_0}^{k_0}k^2\delta \rho_c(k)dk.
\end{eqnarray}
The magnetization of the ground state in the thermodynamic limit is 
\begin{eqnarray}
M_s=\frac 14\lim_{L\to \infty} \left\{ 2 L\int_{-k_0}^{k_0}\rho_L^c(k)dk-
4L\int_{-\Lambda}^{\Lambda}\rho_L^s(\lambda)d\lambda+3 \right\}=0,
\end{eqnarray}
where a prefactor $1/2$ comes from our use of reflection symmetry (cf. above), 
and the last term is included to cancel the contributions 
from the holes at $k=0$ and $\lambda=0$ in the distributions 
$Z_L^c(k)$ and $Z_L^s(\lambda)$, respectively, and that of the impurity.
We therefore have a complete screening of the Kondo impurity for ferromagnetic
exchange, in agreement with the Furusaki-Nagaosa conjecture\cite{4}. Our result
implies that a repulsive boundary potential is not detrimental to the 
ferromagnetic Kondo effect -- at least up to the magnitude considered here.
\par
We now determine the Kondo contribution to the specific heat and the Kondo
temperature. While the thermodynamics of Bethe Ansatz solvable models can be
calculated directly, an alternative, both more practical and more physical,
is provided by exploiting the picture of a Landau-Luttinger liquid put forward
by Carmelo and coworkers\cite{carm}. Here, the densities of states of the 
quasiparticles at the Fermi energy $E_F = 0$ are 
\begin{eqnarray}
N_L^c(0)=\frac{d Z_L^c(k_0)}{d \epsilon_L^c(k_0)}=\frac1{2\pi v_c}
+\frac 1L\frac{\delta \rho_c(k_0)}{\epsilon_c'(k_0)}+O(\frac 1{L^2}),\nonumber\\
N_L^s(0)=\frac{d Z_L^s(\infty)}{d \epsilon_L^s(\infty)}=\frac1{2\pi v_s}
+\frac 1L\frac{\delta \rho_s(\infty)}{\epsilon_s'(\infty)}+O(\frac 1{L^2}),
\end{eqnarray}
where $v_c$ and $v_s$ are the velocities of the charge and spin fluctuations,
respectively. The densities of states directly determine the 
low temperature specific heat and magnetic susceptibility and, using standard
expressions from Fermi liquid theory, we obtain the additional
contributions due to the Kondo effect as
\begin{eqnarray}
\delta C & = & \frac{\pi \delta \rho_c(k_0)}{3Lv_c\rho_c(k_0)}T
+\frac{\pi\delta \rho_s(\infty)}{3Lv_s\rho_s(\infty)}T,\nonumber\\
\delta \chi & = & \frac {\delta \rho_s(\infty)}{L\rho_s(\infty)}\chi_0,
\end{eqnarray}
where $\chi_0$ is the susceptibility of the bulk. Here, we have used the fact
that the finite-size corrections to the densities of states in the absence of 
the Kondo impurity are of $O(1/L^2)$ implying that all the contributions $O(1/L)$
are due to the potential scattering in the charge and the Kondo effect in the
spin sector. The Kondo temperature $T_K$, playing the role of the Fermi 
temperature in the local Fermi liquid generated by the impurity\cite{noz},
can be deduced from the spin part of the
impurity specific heat as
\begin{eqnarray}
T_K=\frac{3}{2}\pi n v_s\frac{\rho_s(\infty)}{\delta \rho_s(\infty)},
\end{eqnarray}
where $n$ is the density of electrons in the system. 

The low-temperature thermodynamics is that of a local Fermi liquid, in agreement
with one of the two alternatives permitted by conformal field theory.\cite{froejdh}
It may perhaps surprise that our results also fall into the framework of the 
Furusaki-Nagaosa analysis of thermodynamics.\cite{4} To see this, notice that
the term responsible for the anomalous scaling of the specific heat in their work
is generated by tunneling (in order $J^{-1}$) across the Kondo impurity. In our
case, however, the open boundary conditions exclude such tunneling processes at
all energy scales. Putting the respective coupling constant to zero will produce
an excess specific heat linear in temperature, as we have found. This also suggests
that the two alternatives given by conformal field theory\cite{froejdh} could, in
fact, be just reflect one generic type of scaling behavior and be connected by 
simply varying this effective tunneling matrix element. 

Our finding of local Fermi liquid thermodynamics is a consequence of  the
Landau-Luttinger framework.\cite{carm} The structure of the Bethe Ansatz 
solution guarantees that this picture can indeed be applied. The anomalous scaling
of the specific heat cannot be found in such a picture. Whether the corresponding
microscopic model can be solved by Bethe Ansatz remains to be seen.
Quite generally, the Landau-Luttinger liquid picture can be used to 
determine the complete changes in the
thermodynamics induced by boundaries in integrable quantum systems. Our above
determination of the excess specific heat is the first successful application
of this idea. 
\par
Case (ii): The ground state is described by $N-2$ real $k$'s and two imaginary
$k$'s at $\pm ic$ and a real set of $\lambda$.
The two imaginary $k$'s correspond two bound states of electrons around the 
impurity. In the thermodynamic limit, we can again define densities for the
\em real \rm roots as in (\ref{dens}) and their finite size corrections as in
(\ref{finite}). They satisfy the set of integral equations
\begin{eqnarray}
\rho_c(k) & = &\frac 1\pi + \int_{-\infty}^\infty 
K \left (k-\lambda,\frac c2 \right)\rho_s(\lambda)d\lambda,\nonumber\\
\rho_s(\lambda)& = & \int_{-k_0}^{k_0}K \left(\lambda-k,\frac c2\right)\rho_c(k)dk-
\int_{-\infty}^\infty K(\lambda-\lambda',c)\rho_s(\lambda')d\lambda',
\nonumber \\
\delta \rho_c(k) & = & - K\left(k,\frac c2\right) - 2K(k,c)-\int_{-\infty}^\infty 
K\left(k-\lambda,\frac c2\right)\delta \rho_s(\lambda)d\lambda,\nonumber\\
\delta \rho_s(\lambda) & = & 
2K\left(\lambda,\frac {3c}2\right)+K(\lambda,c) \nonumber \\
\label{imag}
& + & \int_{-k_0}^{k_0}
K\left(\lambda-k,\frac c2\right)\delta \rho_c(k)dk-
\int_{-\infty}^\infty K(\lambda-\lambda',c)\delta \rho_s(\lambda')d\lambda'.
\end{eqnarray}
The magnetization at zero temperature can be calculated as above
\begin{eqnarray}
\label{magbo}
M_s=\frac 14\lim_{L\to\infty}\left\{2L\int_{-k_0}^{k_0}\rho_L^c(k)dk-4L
\int_{-\Lambda}^\Lambda \rho_L^s(\lambda)d\lambda+7\right\}=1.
\end{eqnarray}
Here, $k_0$ is determined from $\int_{-k_0}^{k_0} \rho_L^c(k) dk = (N- 1/2)/L$ .
We have accounted for the contribution of the two bound electrons in the last 
term of (\ref{magbo}). The ground state now is a spin triplet which apparently 
violates the Furusaki-Nagaosa conjecture. To understand this phenomenon, recall
that the present case corresponds to an attractive potential scattering by the
impurity, in addition to ferromagnetic Kondo exchange.  Two electrons will 
then bind to the impurity at $x \approx 0$ and $x \approx L$, 
and form a spin-$\frac 32$ complex with the impurity spin. However, the Coulomb 
interaction is repulsive and induces an indirect antiferromagnetic exchange 
coupling between the conduction electrons and the $S=3/2$-complex. 
The indirect Kondo coupling between the local composite and the conduction 
electrons then is equivalent to a  spin-$\frac 32$
single impurity Kondo problem in a Luttinger liquid. Our exact results then
indicate that, as in the Fermi liquid, an $S=3/2$-impurity is only partially 
screened\cite{11} in a Luttinger liquid and an effective, unscreened 
spin-1 results. 
\par
The specific heat and the Pauli magnetic susceptibility can also be calculated 
with the procedure as discussed in case (i). The total 
susceptibility is divergent at zero temperature due to the nonvanishing moment 
in the ground state.

The absence of complete Kondo screening and the associated changes in the 
thermodynamics do not fall into the two universality classes found by conformal
field theory.\cite{froejdh} Notice, however, that they are a direct consequence
of the attractive potential scattering from the Kondo impurity included in our
model in the present case. Our results then imply that new universality classes
for the Kondo effect in a Luttinger liquid should exist once potential scattering
is included in the conformal field theory. 
\par
In summary, we have studied the low energy properties of a ferromagnetic Kondo
problem in the 1D $\delta$-potential Fermi gas. The model Hamiltonian is 
exactly solvable
for two special sets of coupling constants. In one case (repulsive potential
scattering off the impurity), the local magnetic moment is completely screened 
in the ground 
state which confirms the Furusaki-Nagaosa conjecture, while in the other case
(attractive potential scattering), the ground state is a spin triplet. 
The nonvanishing moment
for the latter case is attributed to the formation of the local spin-$\frac 32$ 
composite which is partially screened by the Kondo effect due to antiferromagnetic
exchange with the remaining conduction electrons. The
specific heat and the susceptibility follow the laws of the single channel Kondo
system, as a consequence of the 
strong backscattering\cite{12}(open boundary) included in the present model.
To the best of our knowledge, this is the first exact solution of the Kondo
problem in an interacting many-body system.
\par
The authors acknowledge support from SFB 279 of Deutsche Forschungsgemeinschaft.
J. V. is a Heisenberg Fellow of DFG.


\begin{references}
\bibitem{1}F.D.M. Haldane, J. Phys. {\bf C 14}, 2585 (1981).
\bibitem{2}For a review, see, e.g., J. Voit, Rep. Prog. Phys. {\bf 58}, 977 
	(1995).
\bibitem{3}D.-H. Lee and J. Toner, Phys. Rev. Lett. {\bf 69}, 3378 (1992).
\bibitem{4}A. Furusaki and N. Nagaosa, Phys. Rev. Lett. {\bf 72}, 892 (1994).
\bibitem{wang2} Y. Wang, Chin. Phys. Lett., Sept. 1996 (in press).
\bibitem{12}M. Fabrizio and A.O. Gogolin, Phys. Rev. B {\bf 51}, 17827 (1995).
\bibitem{sching} A. Schiller and K. Ingersent, Phys. Rev. B {\bf 51}, 4676 (1995).
\bibitem{froejdh} P. Fr\"{o}jdh and H. Johannesson, Phys. Rev. Lett. {\bf 75},
	300 (1995).
\bibitem{yang} C. N. Yang, Phys. Rev. Lett. {\bf 19}, 1312 (1967).
\bibitem{5}C.L. Kane and M.P.A. Fisher, Phys. Rev. Lett. {\bf 68}, 1220 (1992);
	 Phys. Rev. {\bf B 46}, 15233 (1992).
\bibitem{7}H. Schulz, J. Phys. {\bf C18}, 581 (1985).
\bibitem{8}A. Foerster and M. Karowski, Nucl. Phys. {\bf B 396}, 611 (1993); 
	{\bf 408}, 512 (1993).
\bibitem{wang} Y. Wang, J. Voit, and F.-C. Pu, Phys. Rev. B {\bf 54}, 
	Sept. 15, 1996; cond-mat/9602086.
\bibitem{10}H. Asakawa and M. Suzuki, J. Phys. A: Math. Gen. {\bf 29}, 225 
	(1996).
\bibitem{alca} F. C. Alcaraz, M. N. Barber, M. T. Batchelor, R. J. Baxter, and 
	G. R. W. Quispel, J. Phys. A: Math. Gen. {\bf 20}, 6397 (1987).
\bibitem{6}F. Woynarovich, Phys. Lett. {\bf 108A}, 401 (1985).
\bibitem{9}F. Woynarovich, J. Phys. A: Math. Gen. {\bf 22}, 4243 (1989).
\bibitem{carm} J. Carmelo and A. A. Ovchinnikov, J. Phys.: Condens. Matter {\bf 3},
	757 (1991);
	J. Carmelo, P. Horsch, P. A. Bares, and A. A. Ovchinnikov,
	Phys. Rev. B {\bf 44}, 9967 (1991).
\bibitem{noz} P. Nozi\`{e}res, J. Low Temp. Phys. {\bf 17}, 31 (1974).
\bibitem{11}N. Andrei, K. Furuya and J. Lowenstein, Rev. Mod. Phys. {\bf 55}, 
 	331(1983).



\end{references}
\end{document}